\documentclass[amsmath,amssymb,aps,prl,twocolumn,floatfix,superscriptaddress]{revtex4-2}

\usepackage{graphicx,dcolumn,bm,color,multirow,float}

\begin{document}

\title{Reshaped Weyl fermionic dispersions driven by Coulomb
interactions in MoTe$_2$}

\author{Seoung-Hun Kang}\thanks{These authors contributed equally to this work.}
\affiliation{Korea Institute for Advanced Study, Seoul 02455, Korea}

\author{Sangjun Jeon}\thanks{These authors contributed equally to this work.}
\affiliation{Department of Physics, Chung-Ang University, Seoul 06974, Korea}

\author{Hyun-Jung Kim}
\affiliation{Korea Institute for Advanced Study, Seoul 02455, Korea}
\affiliation{Peter Grünberg Institute and Institute for Advanced Simulation, Forschungszentrum Jülich, Jülich 52428, Germany}

\author{Wonhee Ko}
\affiliation{Samsung Advanced Institute of Technology, Suwon 16678, Korea}
\affiliation{Center for Nanophase Materials Sciences, Oak Ridge National Laboratory, Oak Ridge, Tennessee 37831, United States}

\author{Suyeon Cho}
\affiliation{Department of Chemical Engineering and Materials Science, Graduate Program in System Health Science and Engineering, Ewha Womans University, Seoul 03760, Korea}

\author{Se Hwang Kang}
\affiliation{Department of Energy Science, Sungkyunkwan University, Suwon 16419, Korea}

\author{Sung Wng Kim}
\affiliation{Department of Energy Science, Sungkyunkwan University, Suwon 16419, Korea}

\author{Heejun Yang}
\affiliation{Department of Physics, Korea Advanced Institute of Science and Technology, Daejeon 34141, Korea}

\author{Hyo Won Kim}
\email[E-mail: ]{hyowon98.kim@samsung.com}
\affiliation{Samsung Advanced Institute of Technology, Suwon 16678, Korea}

\author{Young-Woo Son}
\email[E-mail: ]{hand@kias.re.kr}
\affiliation{Korea Institute for Advanced Study, Seoul 02455, Korea}

\date{\today}

\begin{abstract}
We report the direct evidence of impacts of the Coulomb interaction in a prototypical Weyl semimetal, MoTe$_2$, that alter its bare bands in a wide range of energy and momentum. 
Our quasiparticle interference patterns measured using scanning tunneling microscopy are shown to match the joint density of states of quasiparticle energy bands including momentum-dependent self-energy corrections, while electronic energy bands based on the other simpler local approximations of the Coulomb interaction fail to explain neither the correct number of quasiparticle pockets nor shape of their dispersions observed in our spectrum. With this, we predict a transition between type-I and type-II Weyl fermions with doping and resolve its disparate quantum oscillation experiments, thus highlighting the critical roles of Coulomb interactions in layered Weyl semimetals.
\end{abstract}

\maketitle

The proper approximations of electron-electron interactions in solids are pivotal in obtaining their accurate electronic structures1. Especially, the interaction significantly alters band gaps and band dispersions of three-dimensional semiconductors and insulators~\cite{martin}. The layered materials such as graphene and transition metal dichalcogenides (TMDCs) also experience strong electron-electron interaction regardless of their electronic properties. For example, in semi-metallic graphene, its linear bare band dispersion is significantly modified near the Fermi energy~\cite{GONZALEZ1994,Trevisanutto2008,Park2009,Elias2011} and, in semiconducting TMDCs, the quasiparticle energy band gap is enhanced a lot from its mean-field value~\cite{Qiu2013}. Semi-metallic TMDCs are also expected to host non-trivial interaction effects thanks to similar causes in graphene~\cite{Kim2017,Rhodes2017,Song2018,Xu2018,Aryal2019,Hu2020}. Since single-particle energy bands in these semi-metallic materials are also fundamentals in characterizing their topological physics~\cite{Soluyanov2015,Sun2015}, the roles of the long-range interactions in varying energy bands should be examined critically.
  
The orthorhombic MoTe$_2$ (called as T$_d$- or $\gamma$-phase) is a layered TMDC and type-II Weyl semimetals~\cite{Soluyanov2015,Sun2015,Deng2016,Huang2016,Ding2017}. Notwithstanding its semi-metallicity, screening of the interaction is not so effective owing to its van der Waals interaction-bounded layered structure. Some earlier experiments using angle-resolved photoemission spectroscopy (ARPES) and scanning tunneling microscopy (STM), respectively, reported its Weyl points and associated Fermi arcs~\cite{Deng2016,Huang2016,Ding2017}. However, the experimental observation of the Fermi arcs in Weyl semimetals is not so clear due to the concurrence of the bulk electron and hole energy bands in the same energy range~\cite{Deng2016,Huang2016,Ding2017}. The small momentum distance between the pairs of the Weyl points~\cite{Kim2017,Soluyanov2015,Sun2015}, the existence of topologically trivial surface states near the Fermi arc~\cite{Rhodes2017}, and the annihilation of the Weyl point pairs under lattice distortion~\cite{Sie2019,Singh2020} also hinder its clear observation. Moreover, quantum oscillation (QO) measured in recent works does not agree well with computations based on a simple approximation of the Coulomb interactions~\cite{Xu2018,Aryal2019,Hu2020}. Considering recent reports exploring a possible higher-order topological insulating state~\cite{Wang2019,Tang2019} as well as a non-trivial insulating phase~\cite{Jia2020,Wang2021} in MoTe$_2$, it is crucial to clarify effects of electron-electrons interactions and their changes with doping or under different screening environments.

Since the low-energy states involves mostly $d$-orbitals of Mo atom and partly $p$-orbitals of Te atom~\cite{Kim2017}, it is a natural forward step to include the on-site Coulomb repulsion ($U$) for the $d$-orbital~\cite{Kim2017,Aryal2019,Hu2020} within {\it ab initio} methods based on the density functional theory (DFT)~\cite{Hohenberg1964,Kohn1965}. However, the Hubbard $U$ correction within the DFT method (DFT+$U$)~\cite{Dudarev1998} is in short of describing the electronic properties of solids that also have covalent bonds between $d$-orbitals and their neighboring $p$-orbitals~\cite{Leiria2010,Kulik2011,Lee2020,Rubio2020} as does the present case. In addition, the coexistence of electron and hole states with different crystal momentum near the Fermi level ($E_F$) in MoTe$_2$ requires energy-momentum resolved approximations for the screened Coulomb interactions beyond a simple local correction such as the $GW$ approximation~\cite{Hedin1965,Hybertsen1986}.

Here, we identify significant impacts of momentum-dependent self-energy correction on quasiparticle energy dispersions in MoTe$_2$. It alters not only low-energy electronic structures and the associated Weyl points but also high-energy electron and hole bands in the whole Brillouin zone (BZ). This is made possible by comparing quasiparticle interference (QPI) patterns from STM measurements and $GW$ energy band calculations. In a wide range of energy windows around $E_F$, a single-electron and hole pocket are identified, respectively, agreeing well with our QPI measurements, while other methods compute a different number of the pockets and fail to match their shapes. This agreement enables us to predict other quasiparticle properties such as a transition from type-I to type-II Weyl fermions with doping and resolve the dissimilar QO measurements~\cite{Xu2018,Aryal2019,Hu2020}. Therefore, the self-energy corrections in Weyl semimetals are as significant as their roles in semiconductors and is indispensible to consider other topological physics in MoTe$_2$. 

\begin{figure}[t]
\includegraphics[width=1\linewidth]{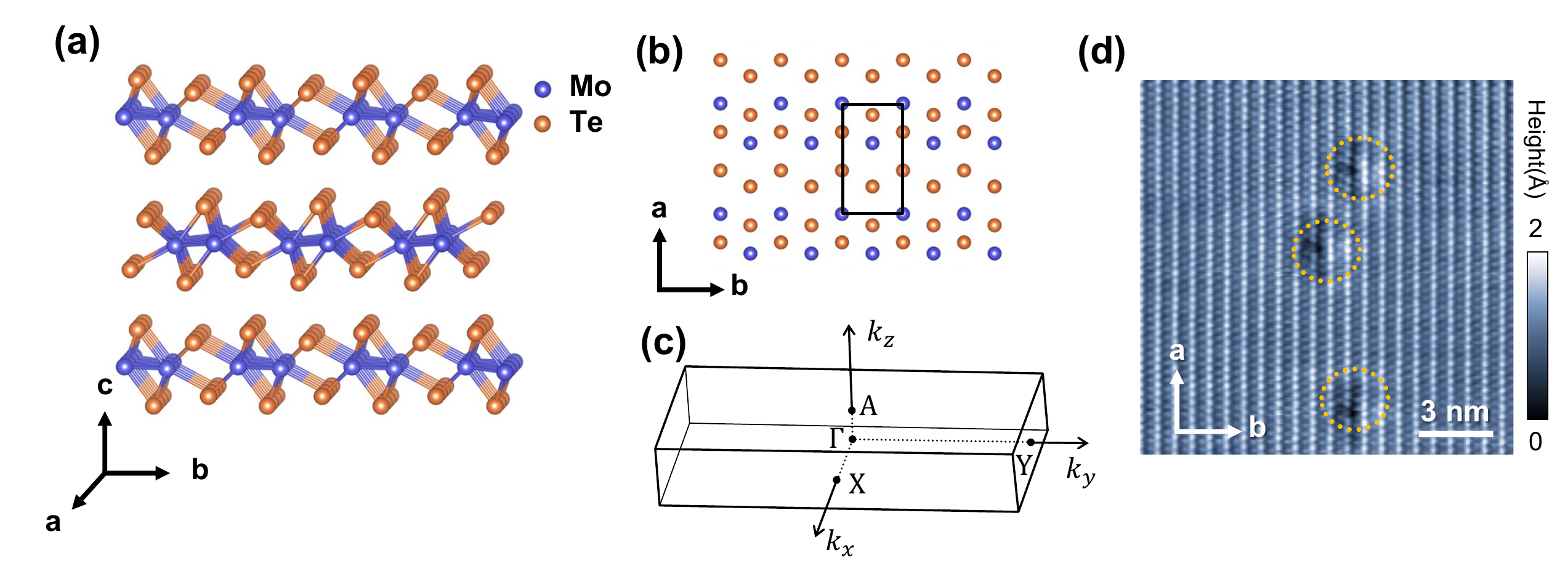}
\caption{Morphology and electronic structure of $\gamma$-MoTe$_2$. The atomic structure of (a) bulk and (b) single-layer, respectively. The black box represents a single-layer unit cell. (c) BZ of orthorhombic MoTe$_2$. (d) STM topography of surface near a defect at V$_s$ = 300~mV (I$_t$ = 1~nA). The dotted yellow circles represent tellurium vacancies. 
\label{Fig1}}
\end{figure}

The bulk Weyl semi-metallic MoTe$_2$ has a distorted 1T’-layer as a unit layer that stacks into an orthorhombic structure at a temperature below 250~K~\cite{Clarke1978}. In each layer, the tellurium layers sandwich the molybdenum layer, as displayed in Fig.~\ref{Fig1}(a). The natural cleavage plane of MoTe$_2$ is the $ab$ plane with a tellurium layer as a surface. The typical topography of the cleaved MoTe$_2$ shows parallel arrays of one-dimensional tellurium atoms along the $a$-axis and 0.6{~\AA} higher than the other parallel arrays of tellurium atoms (see Figs.~\ref{Fig1}(a) to (d)). The tellurium vacancies were observed frequently on a surface, as shown in Fig.~\ref{Fig1}(d). The vacancies in the surface enable us to determine the relationship between energies of elastically scattered electrons and their scattering vectors in STM measurements (See Fig. S1 of Supplemental Material (SM) for details). By Fourier-transforming spatial scanning tunneling spectroscopic images at a given energy, we obtain the multiple dispersion relations between momentum and band energies that can be interpreted as quasiparticle interference (QPI) induced by the vacancies~\cite{Hoffman_2011}. This is equivalent to the joint density of states (JDOS) at a given momentum and its associated constant energy surface~\cite{Hoffman_2011} (Simulation details are given in the SM). Therefore, to understand the experimentally obtained QPI based on theoretically computed JDOS, precise quasiparticle energy bands is a prerequisite. 

\begin{figure}[t]
\includegraphics[width=1\linewidth]{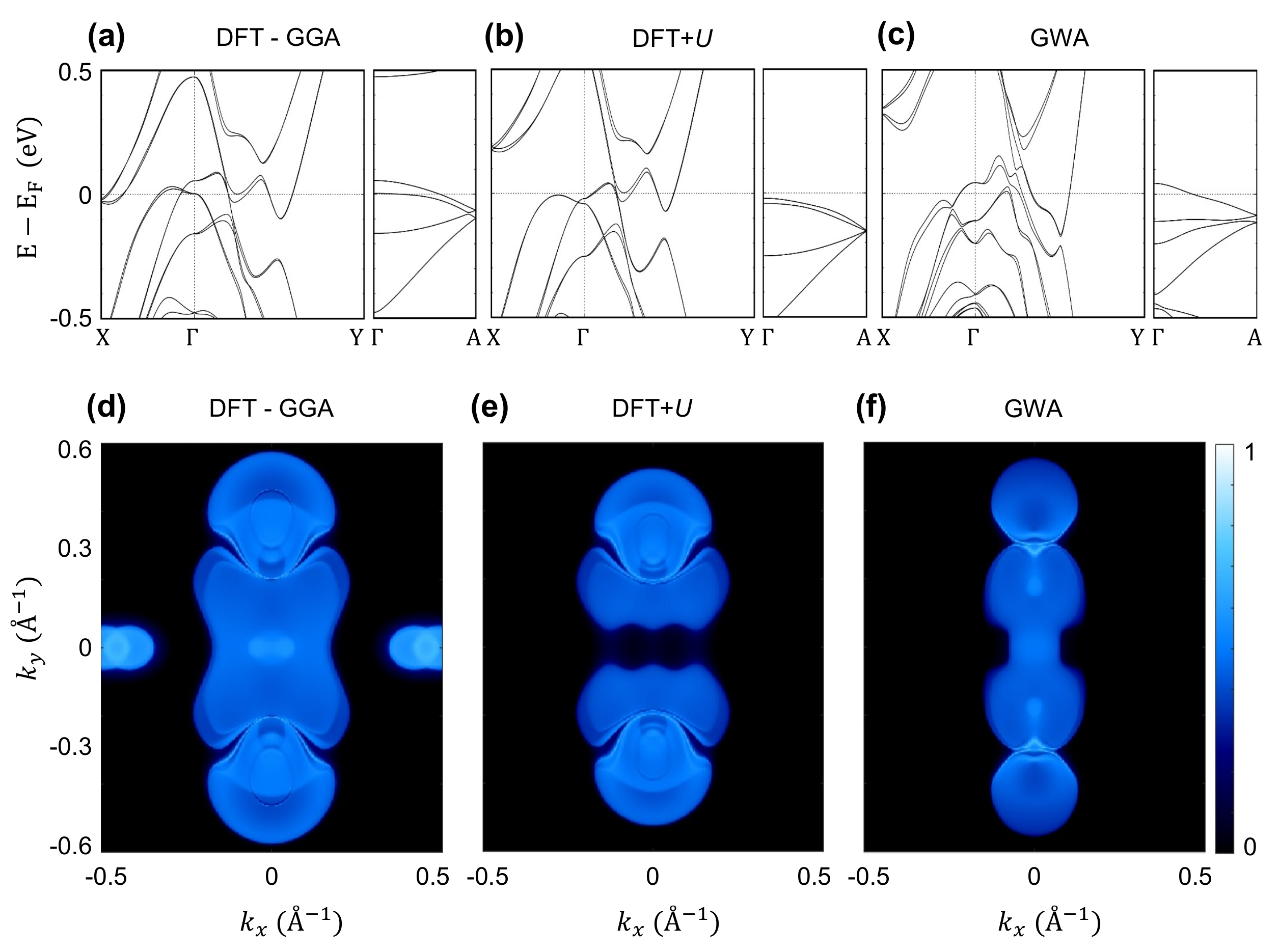}
\caption{Electronic energy bands and projected density of states. (a)-(c) Band structures for orthorhombic MoTe$_2$ using DFT, DFT+$U$ with $U_{eff}$ = 3~eV and the $G_{0}W_{0}$ approximation (GWA), respectively. (d)-(f) The projected density of states on the (001) surface at $E_F$ with the different approximation of the Coulomb interaction. The momentum $k_x$ is along the ${\Gamma}$X direction, and $k_y$ is along the ${\Gamma}$Y direction. 
\label{Fig2}}
\end{figure}

We first compare the electronic energy band structures using DFT with a mean-field type exchange-correlation functional with a generalized gradient correction (DFT-GGA)~\cite{Perdew1996}, DFT+$U$~\cite{Dudarev1998} with $U$ of 3~eV~\cite{Xu2018}, and the $G_{0}W_{0}$ approximation (GWA)~\cite{Hedin1965,Hybertsen1986}, respectively, as shown in Figs.~\ref{Fig2}(a)-(c) (computational parameters are in SM). It is immediately noticeable that the number of electron and hole pockets at the charge-neutral point using different approximations differ from each other. The two superimposed electron pockets exist at the BZ boundary of X-point in the DFT-GGA band while they disappear with the other two methods because of the strong interaction effect near the X-point~\cite{{Kim2017},{Aryal2019}}. We note that the previous ARPES experiments~\cite{Xu2018} also report no such pockets. Besides the difference in the electron pocket's energetic position at the X-point, the overall shape of energy bands from the DFT+$U$ is similar to one from DFT-GGA because the local Hubbard $U$ shifts up the unoccupied DFT-GGA states more or less uniformly. Another difference between them is a shift down of the hole band at the BZ center with $U$ so that the hole pocket at the $\Gamma$-point moves toward the Y-point in DFT+$U$ bands. So, the shape and position of hole pockets in the momentum space will strongly depend on doping within DFT+$U$. 

Energy bands with the GWA in Fig.~\ref{Fig2}(c) markedly differ from band structures using the first two methods. A single-electron pocket along the $\Gamma$-Y direction exists in sharp contrast to double ones with the other methods (here, we neglect spin-orbit split pockets for counting), and a single-hole pocket at BZ center exists, unlike the case of DFT-GGA. This leads to significant changes in the momentum-resolved local density of states at the $E_F$ projected on (001) surface, as shown in Figs.~\ref{Fig2}(d)-(f). The electron pockets near the X-point from the DFT-GGA calculation disappear entirely, and the size of the hole pocket near the $\Gamma$-point becomes smaller, well consistent with ARPES experiments~\cite{Xu2018}. These changes in shape and the number of quasiparticle pockets will play decisive roles in determining QPI patterns.

\begin{figure}[t]
\includegraphics[width=1\linewidth]{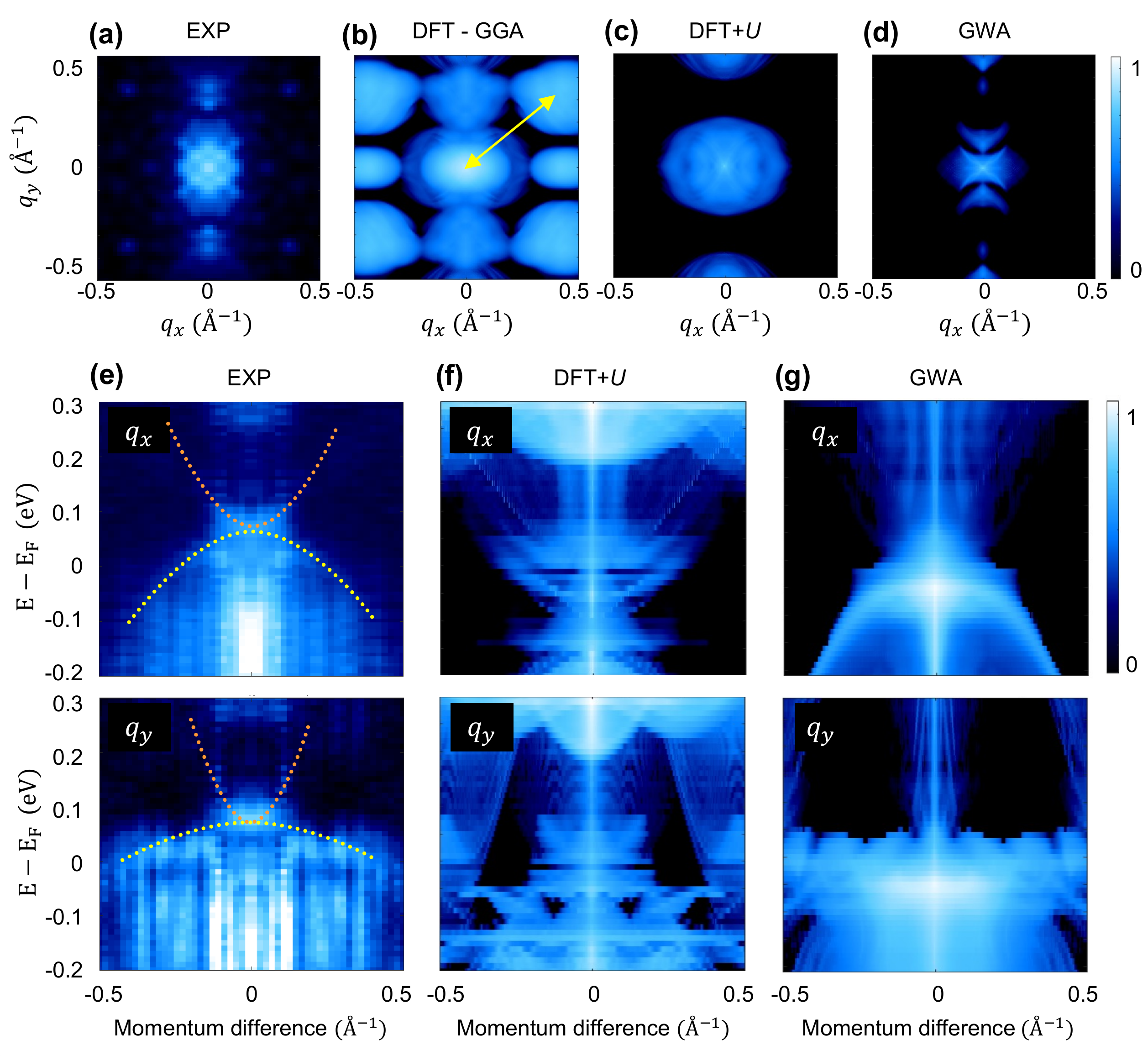}
\caption{Quasiparticle interference patterns on (001) surface of MoTe$_2$. (a) Experiment and (b)-(d) simulated quasiparticle interference (QPI) pattern on the (001) surface at given constant energies. By comparing several constant energy QPIs near the Fermi energy ($E_F$) between experiments and simulations, an experiment QPI at +60~meV is shown in (a) and a QPI pattern with the $G_{0}W_{0}$ approximation (GWA) at 0~meV shown in (d) are found to match each other very well. No QPI patterns with DFT-GGA and DFT+U match experimental data around the $E_F$. QPI patterns at the $E_F$ are displayed for DFT-GGA in (b) and DFT+U in (c), respectively. The vector $q_x$ and $q_y$ are the scattering interference along the $k_x$- and $k_y$-direction, respectively. (e) Measured QPI dispersion along the $q_x$- and $q_y$-direction (top and bottom panels) with energy range from -200 to 300~meV, respectively. Dotted lines are guides for the eye. Simulated QPI dispersions using the DFT+$U$ method in (f) and the GWA in (g). }
\label{Fig3}
\end{figure}

We compute the QPI patterns from JDOSs for given energies and then, consider QPIs in the whole energy-momentum space 
to compare them with our corresponding measurements. Since our measurement conditions cannot be simulated perfectly, the computed QPI patterns with various energies near the charge-neutral point are compared with measurements as best as they can match (for detailed comparisons, see Fig. S2 in SM). We find that the shift of $E_F$ by $-$60~meV with respect to the charge-neutral point in the GWA is appropriate for comparison with measurements as shown in Fig.~\ref{Fig3} and Fig. S2. This shift is also consistent with our previous STM study~\cite{Kim2020} using a similar sample. The QPI simulation shows several distinct interference patterns along the direction connecting the pockets due to multiple quasiparticle pockets. So, careful comparison between simulated and measured patterns can determine the quasiparticle energy bands of the system. 

All three constant energy QPI simulations are quite different from each other, as shown in Fig.~\ref{Fig3}. Among them, the pattern based on the GWA agrees with the measurement mostly. A pair of electron pockets near the $\pm$X-point and $\Gamma$-point in the contour by the DFT-GGA method (Fig.~\ref{Fig2}(d)) induces the corresponding inter-pocket interference signals near $(q_x, q_y) = (\pm0.35\text{\AA}^{-1}, \pm(\mp)0.3\text{\AA}^{-1})$ in the DFT-GGA QPI map, as indicated by the yellow arrow in Fig.~\ref{Fig3}(b). However, our QPI measurement in Fig.~\ref{Fig3}(a) does not show such interferences there. With DFT+$U$ and GWA, they are absent owing to the proper correction of the local Coulomb interaction. Except this, the QPI patterns from the DFT+$U$ calculation and DFT-GGA share two main features; first, multiple linear interference patterns cross the ${\bm{q}}$ = 0 point; second, a pair of large round-shaped patterns surround the crossed patterns. The former crossed interference signals come from intra-pocket scattering, appearing in all simulations as well as our experiment. The latter double-shell-shaped patterns originate from the inter-pocket scatterings. We cannot remove the outer round-shape interference pattern with DFT-GGA and DFT+$U$ within our simulation energy window shown in Figs.~\ref{Fig3}(e)-(g). Moreover, even with such large doping, the constant energy QPI simulation from DFT does not match the measurements at all (Fig. S2). So, we conclude that single-particle band structures from the first two methods do not reflect measured energy bands. 

Our QPI simulation near the charge-neutral point based on the GWA ($GW$-QPI hereafter) has no closed round-shaped pattern encircling the BZ center, unlike the previous two results, agreeing with our measurements. Although there is a slight difference between the sizes of patterns in the $GW$-QPI and the measured one, they share all essential features such as crossed central interferences, satellite interference pattern distributed along the $\Gamma$Y direction, and absence of large round-shaped double-shell interferences as shown in Fig.~\ref{Fig3}(d). The size difference may come from different screening environments between simulation and experiment; our $GW$-QPI is obtained by projecting bulk bands to the (001) plane, thus assuming bulk dielectric screening everywhere while screening in our actual measurement on a doped sample has the vacuum as a half-upper space on the surface. 

Now, the measured QPI maps in the whole phase space are compared with simulations, as shown in Fig.~\ref{Fig3}(e)-(g). Unlike previous studies for limited energy windows~\cite{Deng2016,Ding2017}, we construct energy varying QPI maps in the whole BZ with a broader energy range from –200 to +300~meV. In measured QPI maps, the hole-like quadratic-shaped dispersion patterns (yellow dotted line in Fig.~\ref{Fig3}(e)) spread relatively wider than those above electron-like dispersion patterns (orange dotted line). Moreover, the hole-side map is very anisotropic: the bright outer dispersion along the $q_x$-direction is quite steeper than one along the $q_y$-direction. All aforementioned major features in our experimental QPI map in Fig.~\ref{Fig3}(e) are well captured by the $GW$-QPI map in Fig.~\ref{Fig3}(g). Unlike $GW$-QPI, the QPI maps with DFT+$U$ show very distinct patterns if compared with measurements in both directions, as shown in Fig.~\ref{Fig3}(f). This dramatic difference between the QPI maps from different computational methods also can be noticed for constant energy QPIs with varying energy (See Fig. S2 in SM). 

Our current comparison between simulated and measured QPI maps implies that the properly screened Coulomb self-energies should be included to understand several intriguing low-energy physical properties of MoTe$_2$. With this establishment, we now turn to its low-energy quasiparticle band structures that are critical to its topological properties. 

\begin{figure}[t]
\includegraphics[width=1\linewidth]{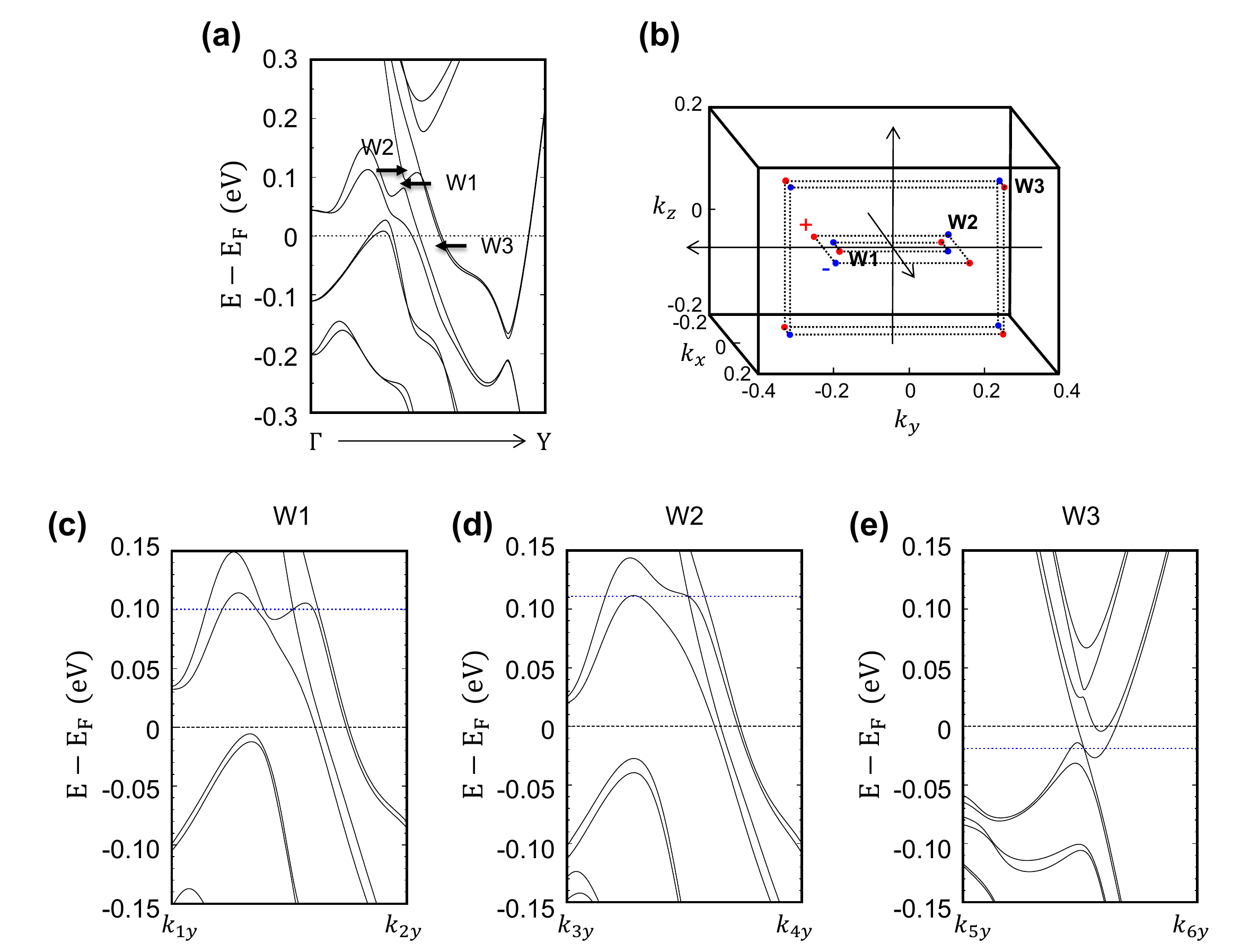}
\caption{Coexistence of type-I and II Weyl points in MoTe$_2$. (a) Electronic band structure obtained with the $G_{0}W_{0}$ approximation along the $\Gamma$-Y direction. Black arrows indicate the Weyl points (WPs). (b) Position of all WPs in the BZ; red/blue dots depict +/- chirality of the WPs. (c)-(e) Magnified band dispersions around Weyl points. Dispersions are drawn along a momentum cut parallel to $k_y$ passing through the W1, W2 and W3, respectively. In abscissa of (c)-(e), $k_{1y}$ (in unit of {\AA}$^{-1}$) is (0.035, 0.020, 0.000), $k_{2y}$ = (0.035, 0.220, 0.000), $k_{3y}$ = (0.059, 0.023, 0.000), $k_{4y}$ = (0.059, 0.223, 0.000), $k_{5y}$ = (0.013, 0.065, 0.429), and $k_{6y}$ = (0.013, 0.265, 0.429), respectively. The dotted black line is $E_F$, and the blue dotted line represents the energy of WPs. 
\label{Fig4}}
\end{figure}

We find that the total number of Weyl points (WPs) and type of Weyl fermionic dispersions are very sensitive to the Coulomb interaction as much as they are to the lattice distortions~\cite{{Kim2017},{Sie2019},{Singh2020}}. Within our DFT+$U$ calculation, the two sets of Weyl points are on the planes at $k_z$ = 0 and $k_z$ $>$ 0, respectively, and all of them are type-II as shown in Fig. S3, agreeing with a previous study~\cite{Singh2020} while disagreeing with the other~\cite{Xu2018}. We note that all WPs can be accessed by electron doping within DFT+$U$. In sharp contrast to DFT+$U$ results, there are 16 WPs with momentum-dependent self-energy corrections in Fig.~\ref{Fig4}. Moreover, there are eight (denoted as W3 in Fig.~\ref{Fig4}) and four (W2) type-II Weyl fermions below and above the charge-neutral points, respectively, while four (W1) type-I Weyl fermions exist in between them. We note that the energetic positions of WPs within DFT+U are around +20$\sim$30~meV with respect to $E_F$, while those with the GWA spread from –20~meV to +110~meV. We can infer from the quasiparticle dispersions that within realistic doping concentration, one can easily access the type-II Weyl fermions and with slightly more electron doping, the dispersion becomes to be the type-I. With electron doping, one may measure distinct electronic and topological properties owing to the presence of the different types~\cite{Zhang2018}. We also note a very close proximity between type-I and type-II Weyl points. So, a slight change of symmetry by external stimuli will increase the chance to induce a higher-order topological insulating state~\cite{Choi2020}.

Before concluding, we discuss the roles of other important physical parameters in MoTe$_2$ that can change its low-energy quasiparticle dispersions in conjunction with the aforementioned effects of Coulomb interactions. Since its electronic structure is intimately intertwined with its atomic structure, a minute variation of its structure can alter its low-energy bands a lot~\cite{{Kim2017},{Sie2019},{Singh2020}}. This interesting interplay could be useful in realizing desired functionalities if on-target controls are achieved. From a different point of view, however, it is an unfavorable situation to compare disparate experimental results using different samples properly because minute structural variations in each sample owing to many extrinsic causes~\cite{Kim2017} may obscure measured signals. Therefore, it is desirable to consider mutual influences between those factors. 

One notable example related to these aspects is quantum oscillation (QO) measurement. Recent QO measurements and computational results reported so far~\cite{{Rhodes2017},{Hu2020}} show different characteristics of its low-energy Fermi surfaces. We find that the QO simulations also show critical variations depending on various factors such as the degree of approximations for the Coulomb interaction, structural parameters, and doping levels (computational details in the SM). With DFT+$U$, the simulated QOs display strong variations of high-frequency signals depending on doping level, as shown in Fig. S4. A simulated QO without doping in Fig. S4 does not seem to match measurements, unlike a previous study~\cite{Hu2020} owing to the minute structural differences between the two cases as shown in Fig. S5. With GWA, however, the minute structural changes do not alter the simulated QO abruptly (not shown here), unlike the cases with DFT+$U$. With GWA, we find that the simulated QOs show gradually decreasing (increasing) high-frequency signals as hole (electron) doping increases, respectively as shown in Fig. S4, indicating that the disparate QO measurements reflect different doping conditions. 

We have studied the effects of the Coulomb interactions on the low-energy electronic properties of a prototypical Weyl semimetal, MoTe$_2$. Precise comparisons between experimental and theoretical quasiparticle bands have been achieved over the wide range of energy in the vicinity of the $E_F$. From these observations, we can discern the appropriate theoretical approach to deal with the strongly screened Coulomb interaction from insufficient ones, thus establishing the roles of the interaction in reshaping the whole landscape of low-energy bands in the three-dimensional layered Weyl semi-metallic materials. 

\begin{acknowledgments}
Y.-W.S. was supported by the National Research Foundation of Korea (NRF) (Grant No. 2017R1A5A1014862, SRC program: vdWMRC center) and KIAS individual Grant No. (CG031509). S. Jeon was supported by the NRF (G2020R1A5A1016518 \& 2020R1F1A1067268) and Samsung Advanced Institute of Technology (SAIT). S.C. was supported by the Basic Science Research Program of the NRF (2020R1A2C2003377). The experiment data analysis was conducted at the Center for Nanophase Materials Sciences, which is a DOE Office of Science User Facility. We thank the CAC of KIAS for providing computing  for this work. 
\end{acknowledgments}

%

\onecolumngrid
\clearpage
\begin{center}
\textbf{\large Supplemental Material for ``Coulomb interaction-driven reshaped Weyl fermionic dispersions in MoTe$_2$''}
\end{center}
\setcounter{equation}{0}
\setcounter{figure}{0}
\setcounter{table}{0}
\setcounter{page}{1}
\setcounter{section}{0}
\setcounter{subsection}{0}

\maketitle
\makeatletter
\renewcommand{\thesection}{\arabic{section}}
\renewcommand{\thesubsection}{\thesection.\arabic{subsection}}
\renewcommand{\thesubsubsection}{\thesubsection.\arabic{subsubsection}}
\renewcommand{\theequation}{S\arabic{equation}}
\renewcommand{\thefigure}{S\arabic{figure}}
\renewcommand{\thetable}{S\arabic{table}}

\subsection{SAMPLE PREPARATION}
\label{SAMPLE PREPARATION}

High-quality 1T’-MoTe$_2$ single-crystals were synthesized using the NaCl-Flux method. A stoichiometric mixture of Mo and Te powders was sintered with sodium chloride (NaCl) at 1373~K for 30 hours in evacuated silica tubes. Then, the samples were cooled to 1223~K at a rate of 0.5~K/hr, followed by rapid cooling to room temperature. The resulting 1T’-MoTe$_2$ single-crystal exhibited large magnetoresistance and residual resistance ratio, exceeding 32,000 and 350, respectively~\cite{{Keum2015},{Cho_2017}}. 

Scanning tunneling microscopy/spectroscopy measurements 
We performed the STM/STS experiments in a commercial low-temperature STM (UNISOKU Co., Ltd., Osaka, Japan) at 2.8~K. A 1T’-MoTe$_2$ single-crystal sample was cleaved in an ultrahigh vacuum chamber (~10$^{-10}$ Torr) at room temperature and then transferred to the low-temperature STM sample stage, where the temperature was maintained at 2.8~K. The STS measurements were performed using a standard lock-in technique with a bias modulation of 5~mV at 1~kHz.

\subsection{FIRST-PRINCIPLES CALCULATIONS}
\label{FIRST-PRINCIPLES CALCULATIONS}
We performed ab $initio$ calculations based on density functional theory (DFT)~\cite{{Hohenberg1964},{Kohn1965}} as implemented in the Vienna ab initio simulation package (VASP)~\cite{{Kresse1993},{Kresse1996}} with projector augmented wave potentials~\cite{{PAW1994},{Kresse1999}} and spin-orbit coupling. The Perdew-Burke-Ernzerhof (PBE) form~\cite{Perdew1996} was employed for the exchange-correlation functional with the generalized gradient approximation (GGA). For the DFT+$U$ calculations, the Dudarev scheme~\cite{Dudarev1998} was used. The constant Hubbard repulsion of $U$ = 3~eV acting on the Mo 4$d$ states was used following the previous studies~\cite{Xu2018}. The self-energy correction is obtained through a one-shot calculation of Green’s function and screened Coulomb interaction, i.e., $G_{0}W_{0}$ approximation (GWA)~\cite{{Hedin1965},{Hybertsen1986}}. The numbers of occupied and unoccupied bands are 96 and 264, respectively. The Wannier interpolation procedure was performed using the WANNIER90 code~\cite{MOSTOFI2014}. The energy cutoff was set to 450~eV for all calculations with the experimental crystal structure\cite{Qi2016}. The BZ was sampled using a $6\times14\times4$ $\Gamma$-centered k-grid in DFT and DFT+$U$, and $4\times12\times4$ $\Gamma$-centered k-grid in the GWA, respectively.

\begin{figure}[p]
\includegraphics[width=0.5\columnwidth]{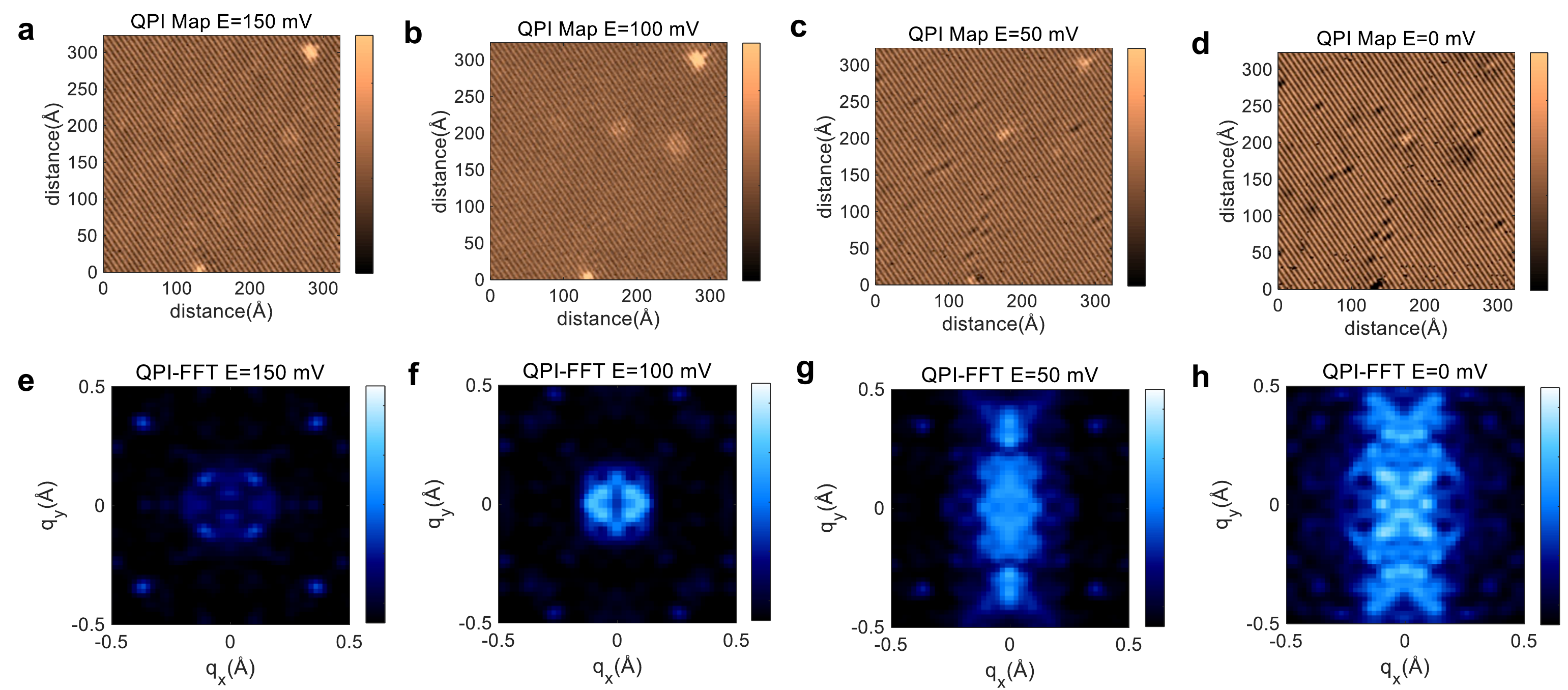}
\caption{STM conductance maps and their QPI patterns. a-d STM conductance maps for various energies ($V_set$ = 500~mV, $V_osc$ = 5~mV, $I_set$ = 0.5~nA). e-f The QPI patterns directly Fourier transformed the conductance maps at the given energies. QPI patterns are rotated 55.7 degrees to match the $q_x$- and $q_y$-directions along the high symmetric momentum axes and symmetrized based on the crystal symmetry of the MoTe$_2$. The sharp peaks that appeared near the corner of the QPI patterns are originated from the 60~Hz background noise in the conductance maps.
\label{FigS1}}
\end{figure}

\begin{figure}[p]
\includegraphics[width=0.5\columnwidth]{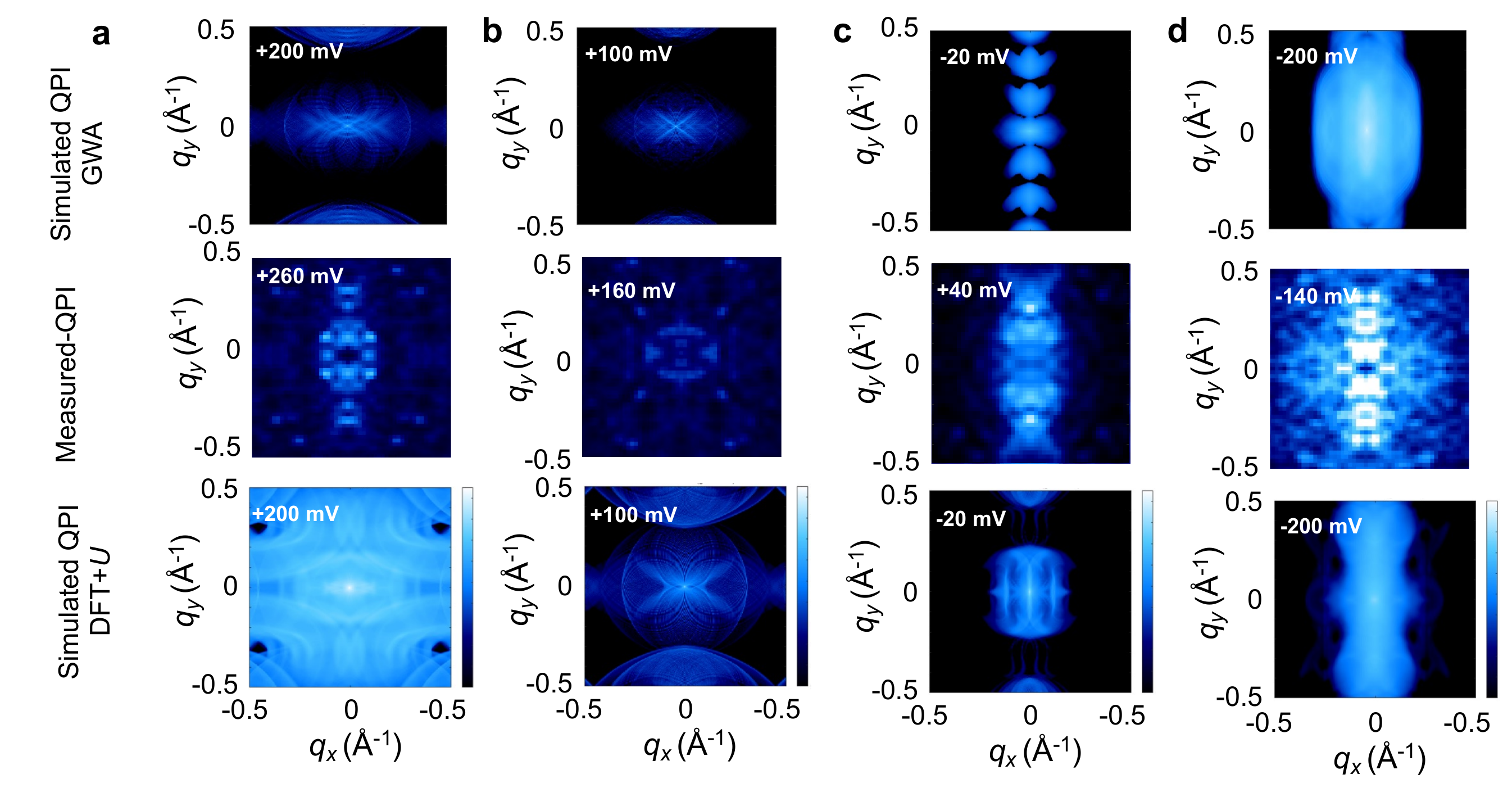}
\caption{QPI pattern at high energy levels. From left to right columns, simulated QPI patterns with $G_{0}W_{0}$ approximation, ones with DFT+$U$ (3~eV), and experimental QPIs are displayed. a, Simulated QPI patterns at +200~mV (left and right panels) and experimentally measured QPI signal at +260~mV (middle panel). Simulated QPIs with different energies are shown in b, E = +100~mV, c -20~mV and d -200~mV, respectively. Experimental QPIs with different energies are in b, E = +160~mV, c +40~mV and d -140~mV, respectively. Considering that the sample is hole-doped, we compared the simulations of which energies are shifted by -60~meV with corresponding measurements.
\label{FigS2}}
\end{figure}

\begin{figure}[p]
\includegraphics[width=0.3\columnwidth]{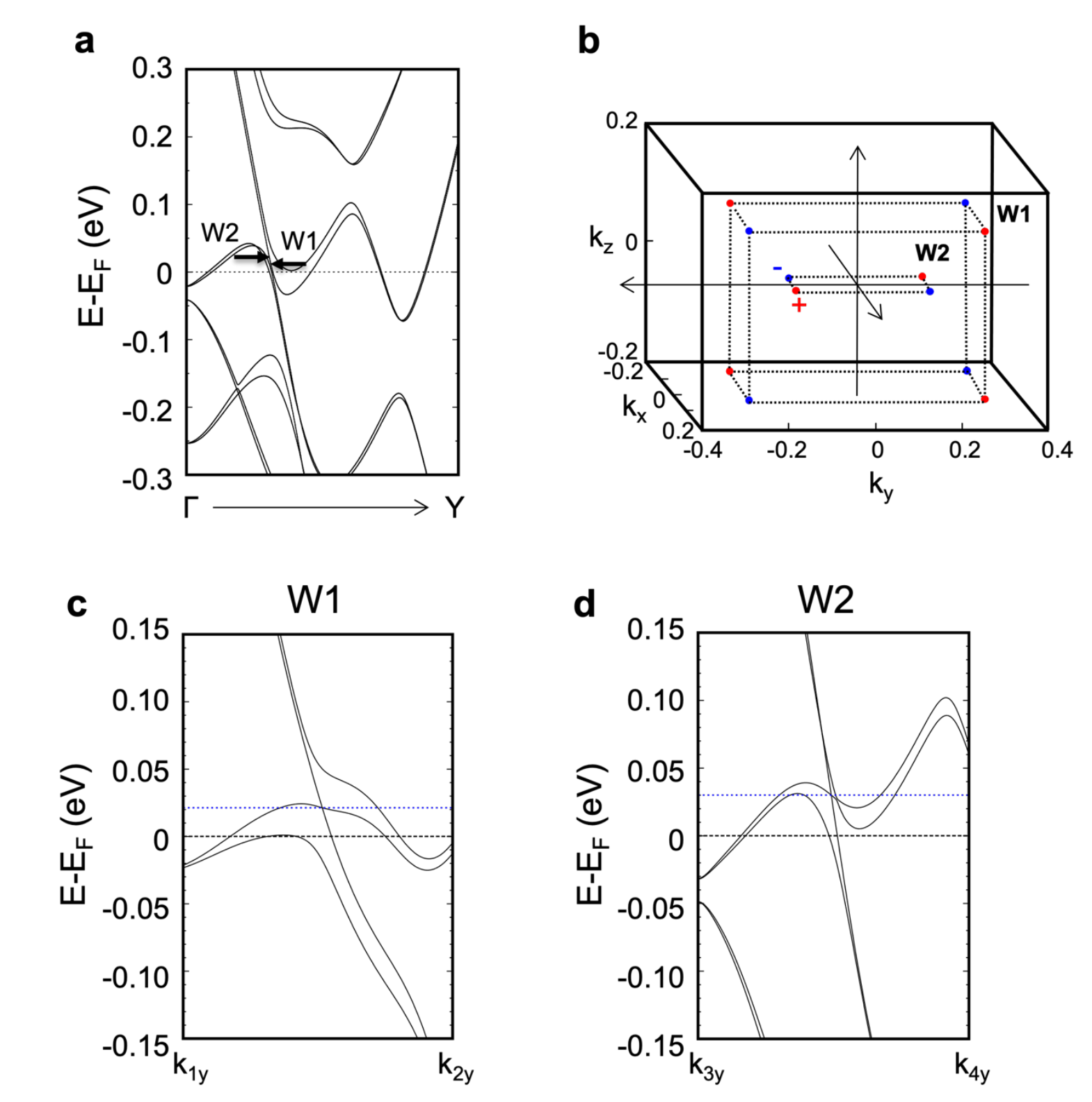}
\caption{Weyl points with DFT+$U$. a Electronic band structure of orthorhombic MoTe$_2$ along the $\Gamma$-Y direction. Black arrows indicate the Weyl points. b Position of all WPs in the BZ; red/blue dots depict +/- chirality of the WPs. c, d Band dispersions along with a momentum cut parallel to $k_y$ passing through the W1, and W2, respectively. The dotted black line is $E_F$, and the blue dotted line represents the energy of WPs. $k_{1y}$ (in unit of {\AA}$^{-1}$) is (0.132, 0.023, 0.200), $k_{2y}$ is (0.132, 0.223, 0.200), $k_{3y}$ is (0.048, 0.000, 0.000), and $k_{4y}$ is (0.048, 0.195, 0.000), respectively. The calculated Weyl dispersions and a number of Weyl points agree with a previous study~\cite{Singh2020}.
\label{FigS3}}
\end{figure}

\subsection{QUASIPARTICLE INTERFERENCE AND QUANTUM OSCILLATION SIMULATIONS}
\label{QUASIPARTICLE INTERFERENCE AND QUANTUM OSCIllATION SIMULATIONS}
From the DFT, DFT+$U$, and GWA results, we obtain the maximally localized Wannier functions for 5$s$- and 4$d$-orbitals of Mo atom and 5$p$-orbitals on Te atoms by using the WANNIER90 code~\cite{MOSTOFI2014} These were used to analyze the surface density of states. The surface projected local density of states and QPI spectra were calculated by the WANNIERTOOLS~\cite{WU2018}, which is based on the iterative Green’s function technique~\cite{Sancho1985}. The spin-dependent QPI map or JDOS, $J_{s}({\bm{q}},\epsilon) = \frac{1}{2}\sum{{\bm{_k}}}\sum_{i=0,\cdots,3}\rho{_i}({\bm{k}},\epsilon)\rho{_i}({\bm{k+q}},\epsilon)$ was calculated with $\rho{_i}({\bm{k}},\epsilon) = -\frac{1}{\pi}Im\{Tr[\sigma{_i}G{_s}({\bm{k}},\epsilon)]\}$. Here, $\sigma{_i}$ ($i$ = 1,2,3) is the Pauli matrix, $\sigma{_0}$ an identity, and $G{_s}({\bm{k}},\epsilon)$ the Green’s function at momentum (${\bm{k}}$) and energy ($\epsilon$) relative to the $E_F$. The angular dependence of the quantum oscillation (QO) frequencies was extracted from the band structure of first-principles calculations using the Supercell K-space External Area Finder (SKEAF) code~\cite{Rourke2012}. The QO was evaluated with the extremal Fermi surface cross-sectional area perpendicular to the magnetic field direction~\cite{shoenberg1984}\: $F$ = ($\frac{\hbar}{2\pi e}$)A where $F$ is the de Haas-van Alphen frequency, $\hbar$ the Planck constant, A is the extremal area and e is the elementary charge.

\begin{figure}[p]
\includegraphics[width=0.5\columnwidth]{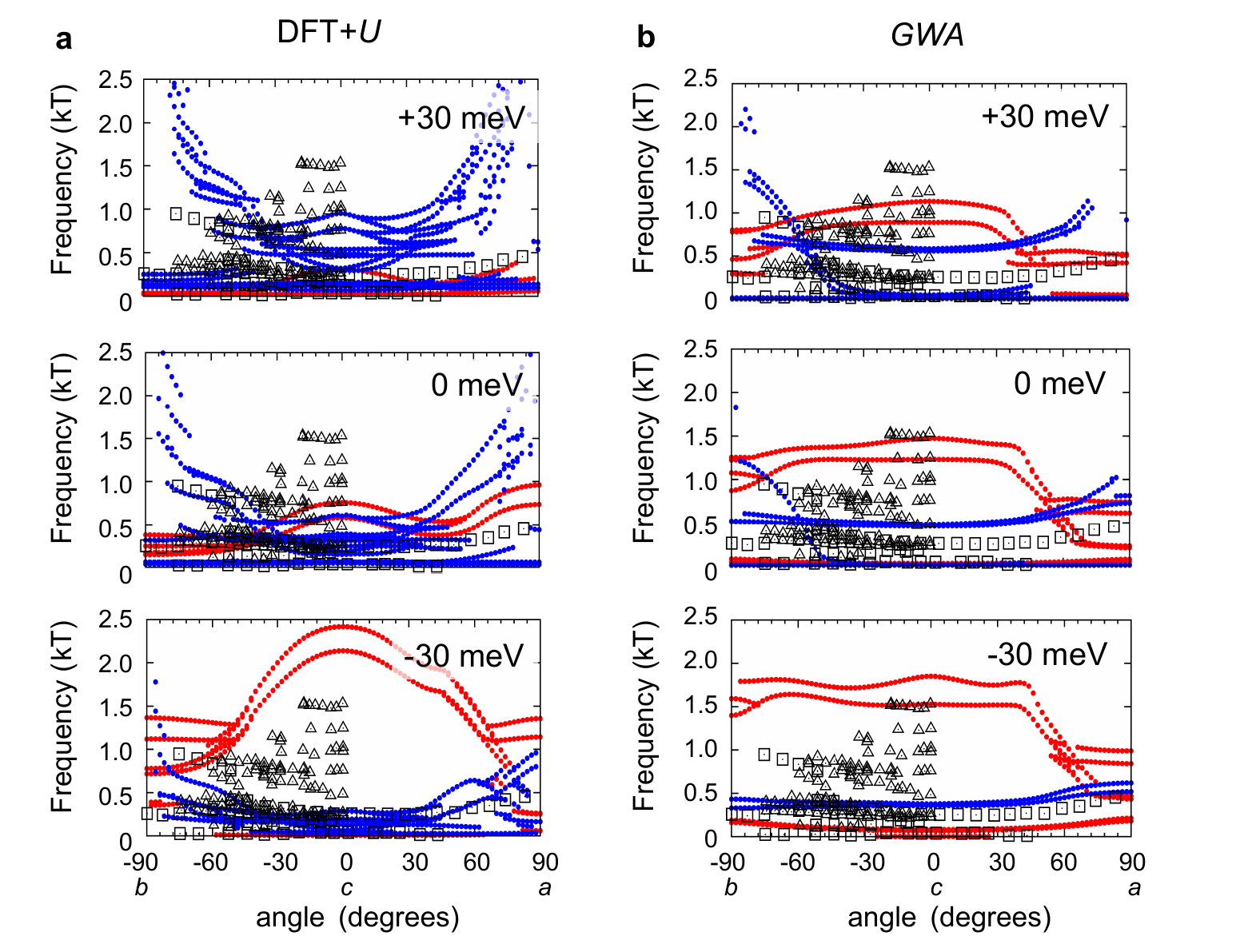}
\caption{The doping effect of quantum oscillation in orthorhombic MoTe$_2$. a Comparison between experimental angular dependence, and the simulated angular dependence of QO in pristine (middle), electron (top) and hole (bottom) doped orthorhombic MoTe$_2$ with DFT+$U$ and b $G_{0}W_{0}$. The angle 0$^{\circ}$ represents the case of $B$  $\parallel$ $c$-axis, -90$^{\circ}$ for $B$ $\parallel$ $b$-axis and +90$^{\circ}$ for $B$ $\parallel$ $a$-axis, respectively. The doping effect of electrons and holes was considered by the rigid shift of $E_F$. The solid red and blue dots depict the hole and electron states for given Fermi energy shifts, respectively. The experimental results were taken from Ref.~\cite{Rhodes2017} (rectangles) and Ref.~\cite{Hu2020} (triangles), respectively.
\label{FigS4}}
\end{figure}

\begin{figure}[p]
\includegraphics[width=0.6\columnwidth]{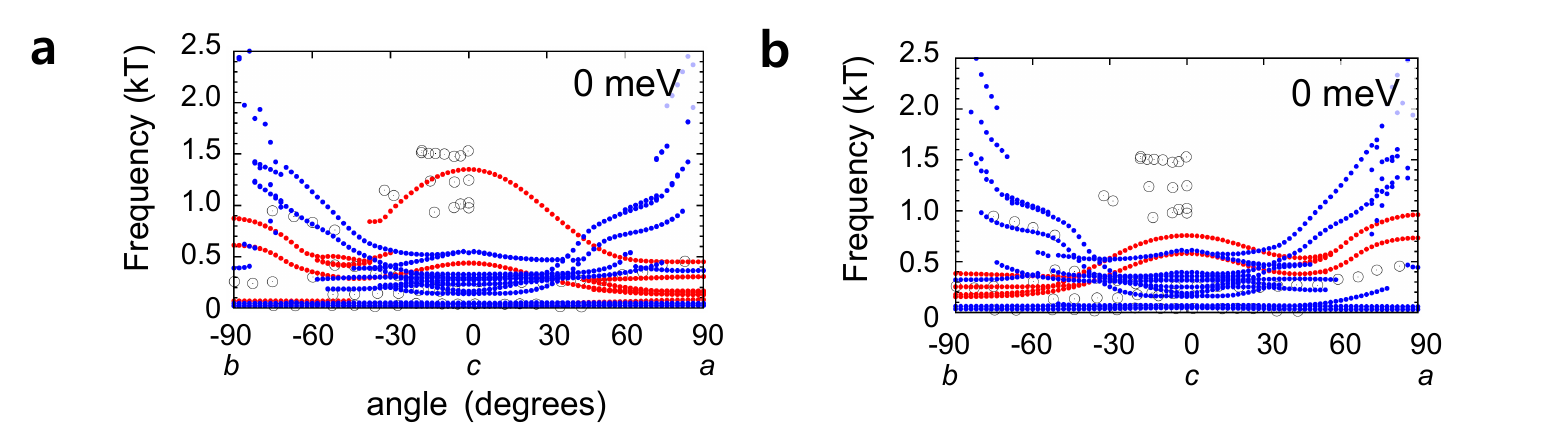}
\caption{The structural effect on quantum oscillation within DFT+$U$. Quantum oscillation simulation results for charge-neutral orthorhombic MoTe$_2$ with a very minute structural difference. a, QO with the experimental lattice constants in Ref. S4, $a$ = 6.310~{\AA}, $b$ = 3.470~{\AA}, and $c$ = 13.930~{\AA}. b QO with the other experimental lattice constants in Ref. S5. The lattice parameters are $a$ = 6.335~{\AA}, $b$ = 3.477~{\AA}, and $c$ = 13.889~{\AA}. Those two are almost similar. The critical difference between the two simulations is a small suppression (0.3\%) of the $c$-axis. The experimental results were taken from Ref.~\cite{Heikes2018} (circles).
\label{FigS5}}
\end{figure}

\end{document}